\documentclass[conference]{IEEEtran}
\IEEEoverridecommandlockouts
\usepackage{amsmath,amssymb,amsfonts}
\usepackage{cite}
\usepackage{algorithmic}
\usepackage{graphicx}
\usepackage{textcomp}
\usepackage{url}
\usepackage{cite}
\usepackage{xcolor}
\usepackage{cleveref}
\usepackage{flushend}
\usepackage{footnote}
\def\BibTeX{{\rm B\kern-.05em{\sc i\kern-.025em b}\kern-.08em
    T\kern-.1667em\lower.7ex\hbox{E}\kern-.125emX}}
\usepackage{fancyhdr}

\fancypagestyle{specialfooter}{%
  \fancyhf{}
  
  \fancyfoot[L]{\small{\textcopyright 2019 IEEE.  Personal use of this material is permitted.  Permission from IEEE must be obtained for all other uses, in any current or future media, including reprinting/republishing this material for advertising or promotional purposes, creating new collective works, for resale or redistribution to servers or lists, or reuse of any copyrighted component of this work in other works.}}
}

\begin{document}

\title{Automatically Extracting Subroutine Summary Descriptions from Unstructured Comments\\
}

\author{\IEEEauthorblockN{Zachary Eberhart, Alexander LeClair, and Collin McMillan}
\IEEEauthorblockA{\textit{Department of Computer Science} \\
\textit{University of Notre Dame}\\
Notre Dame, IN, USA \\
\{zeberhar, aleclair, cmc\}@nd.edu}
}

\maketitle

\thispagestyle{specialfooter}

\begin{abstract}
Summary descriptions of subroutines are short (usually one-sentence) natural language explanations of a subroutine's behavior and purpose in a program.  These summaries are ubiquitous in documentation, and many tools such as JavaDocs and Doxygen generate documentation built around them.  And yet, extracting summaries from unstructured source code repositories remains a difficult research problem -- it is very difficult to generate clean structured documentation unless the summaries are annotated by programmers.  This becomes a problem in large repositories of legacy code, since it is cost prohibitive to retroactively annotate summaries in dozens or hundreds of old programs.  Likewise, it is a problem for creators of automatic documentation generation algorithms, since these algorithms usually must learn from large annotated datasets, which do not exist for many programming languages.  In this paper, we present a semi-automated approach via crowdsourcing and a fully-automated approach for annotating summaries from unstructured code comments.  We present experiments validating the approaches, and provide recommendations and cost estimates for automatically annotating large repositories.
\end{abstract}


\section{Introduction}
\label{sec:intro}

Summary descriptions of subroutines are short (usually one-sentence) natural language explanations of a subroutine's behavior and purpose in a program~\cite{haiduc2010supporting, mcburney2016automatic}.  These summaries are nearly ubiquitous in software documentation, such as in the headers for methods in JavaDocs~\cite{kramer1999api} or Doxygen~\cite{DoxygenWebsite}.  Different studies have for decades verified a conclusion that summaries of source code, and subroutines in particular, provide important clues for programmers at several stages of program comprehension~\cite{von1995program, letovsky1987cognitive, cornelissen2009systematic, maletic2001supporting}.  In some cases, a summary is considered the only trustworthy part of documentation~\cite{roehm2012professional}, since relatively little text (e.g. ``computes a dot product in parallel'') can give a large insight into the code's behavior.

There are two strategies employed to obtain summary descriptions: First and foremost is specially-formatted metadata written by programmers.  This strategy is used by documentation tools such as JavaDocs that automate generation of e.g. HTML pages for viewing documentation, but has the disadvantage of leaving almost all the effort of writing the summaries to human programmers.  Therefore, a second strategy is to automatically generate the summaries based on patterns learned from big data input.  This second strategy saves significant human effort, but relies on large numbers (on the order of millions~\cite{leclair2018funcom}) of high-quality example summaries for learning.  These examples are usually extracted from metadata within large code repositories, but suitable metadata is scarce. LeClair~\emph{et al.}~\cite{leclair2018funcomdata} report only around 5\% of Java methods with suitable metadata, and even fewer suitable for training automatic comment generators.  In short, even automated solutions rely on a large-scale human effort to generate summary descriptions.

There does exist a large, untapped resource of summary descriptions in the form of unstructured header comments found in source code.  Unstructured comments are much more numerous than the well-structured ones in metadata (over 3x as many in one dataset~\cite{leclair2018funcom}), but are much longer and more expansive in scope than short summary descriptions. As we show later in this paper, these comments nearly always have a short summary description embedded in them, but the summary may occur in many locations: surrounded by different text, commented-out code, or even diagrams or logos as ASCII art. This situation is especially prominent for languages without an accepted standard for documentation (e.g. no accepted standard in C versus JavaDocs in Java).  A result is that a vast majority of research focuses on environments such as Java or Python in which it is relatively easy to extract examples, while overlooking more difficult environments such as C or C++. 

Therefore, the challenge is to extract these succinct summary descriptions from unstructured source code comments as a resource for downstream tasks e.g. to use as training examples for automatic summary generation. Manually extracting a summary description from any one function comment is generally not complicated.  Very often the summary is obvious to even a non-expert reader, since the task often does not involve interpreting programming concepts, only filtering obviously unrelated material.  The issue is volume:  hiring programmers to annotate a few hundred examples is feasible.  Hiring them to annotate millions is not.  Thus a question arises as to whether non-experts could be enlisted to perform annotation at a lower cost than experts, or if an automated ``keyphrase'' extraction algorithm may be adapted instead.

In this paper, we propose automated and semi-automated approaches for extracting summaries of subroutines from unstructured comments.  Our work has three components: 1) We hire professional programmers to annotate 1000 summary descriptions in C function comments.  These annotations are extremely expensive, but form a kernel of high-quality annotations around which we can verify other approaches. 2) We hire non-expert workers via Amazon Mechanical Turk~\cite{MTURK} to annotate a further 120000 summary descriptions.  These are less trustworthy than the expert-annotated summaries, but are far more numerous and less expensive.  Finally, 3) we design an automated approach inspired by NLP literature on keyphrase detection.  We train the approach using the 120000 non-expert annotations and test against the 1000 programmer-annotated examples.  We found quite strong performance of the automated and semi-automated approaches in our experiments.

\textbf{Problem Statement}  The problem we target in this paper is the extraction of summary descriptions from unstructured subroutine comments.  By ``summary descriptions,'' we mean a short natural language explanation of code behavior or purpose (maximum 12 words, in line with related work~\cite{leclair2018funcom}, which found that most summary descriptions consisted of fewer than 13 word tokens).  By ``unstructured subroutine comments,'' we mean the long comments immediately preceding methods in source code.  Consider the examples below paired with an ID from our downloadable database (Section~\ref{sec:reproducibility}):


\vspace{0.1cm}

{\scriptsize
\begin{itemize}
\item{1641982:} \texttt{@brief \textbf{Callback every time the parent window is}
\textbf{closed} (pop from the active window stack)}

\vspace{0.1cm}

\item{5354964:} \texttt{VarR8FromUI4 (OLEAUT32.219) \textbf{Convert a VT\_UI4} \textbf{to a VT\_R8.} PARAMS  ulIn [I] Source  pDblOut [O] Destination
RETURNS  Success: S\_OK.  Failure: E\_INVALIDARG, if the 
source value is invalid  DISP\_E\_OVERFLOW, if the value 
will not fit in the destination  DISP\_E\_TYPEMISMATCH, if 
the type cannot be converted}

\vspace{0.1cm}



\item{2997854:} \texttt{ADHOC pattern checking  Now \textbf{look for specific}
\textbf{specific sequences that are easy to optimize.}  Many of these sequences are
characteristic of the compiler  (i.e. it'd probably be a
waste of time to apply these adhoc checks to hand 
written assembly.)}

\item{4442858:} \texttt{XXX: use host strnlen if available ?}

\end{itemize}
}

The summary annotations by professional programmers are in bold.  In the first example, a keyword (``\texttt{@brief}'') is used to indicate the summary.  In the second example, a project-specific format is used.  In the third example, the summary is buried among other text. And in the final example, there is no appropriate summary description at all.  The problem is not that any one summary is difficult annotate; the problem is the volume of different conventions and specific situations for which it is not feasible to manually craft heuristics.

\textbf{Practical Applications}  Our work has two key applications.  First, the model we create can be used to automatically extract summary comments from existing code, to create neatly-readable documentation for code that does not have specially-formatted metadata -- in other words, Doxygen or JavaDoc-like documentation without the extra effort of reformatting a program's comments to fit a specified metadata format.  This application is especially useful for large repositories of legacy code such as the industrial situation described by McMillan~\emph{et al.}~\cite{McMillan:2012:DSS:2337223.2337267}.  A second application is in generating large datasets of code-comment pairs to serve as training data for automatic code summarization tools such as described by LeClair~\emph{et al.}~\cite{leclair2018funcom} and Hu~\emph{et al.}~\cite{hu2018deep}.  These code summarization tools could reach a much wider audience (e.g. C instead of only Java) if large datasets were available.


\section{Background and Related Work}
\label{sec:related}

This section covers the key areas of related work and background technologies from both the Software Engineering and Natural Language Processing research areas.  Mining unstructured data has a long history in SE research, as attested by several topic-focused workshops and surveys~\cite{bavota2016mining, bettenburg2010workshop}; at a high level, this paper fits into the tradition of extracting meaningful snippets of information from unstructured data.

\subsection{Source Code Summarization}

The term ``source code summarization'' was coined around 2009 by Haiduc~\emph{et al.}~\cite{haiduc2010use, haiduc2010supporting} to describe the task of generating short descriptions of code.  Referring to the process as ``summarization'' alludes to a history of work in Natural Language Processing of extractive summarization of documents -- early attempts at code summarization involved choosing a set of \emph{n} important words from code~\cite{hill2009automatically, rodeghero2015eye} and then converting those words into complete sentences by placing them into sentence templates~\cite{sridhara2010towards, rastkar2011generating, sridhara2011automatically, mcburney2016automatic}.  A 2016 survey~\cite{nazar2016summarizing} highlights these approaches around the time that a vast majority of code summarization techniques began to be based on neural networks trained from big data input~\cite{iyer2016summarizing, jiang2017automatically, hu2018deep, leclair2018funcom, alon2018code2seq, fernandes2018structured}.  These NN-based approaches have proliferated, but suffer an Achilles' heel of reliance on very large, clean datasets of examples of code comments. This paper aims to reduce the effects of that weakness by vastly increasing the amount of available data.  In addition, our work differs from these approaches by enabling the extraction of summaries from existing unstructured comments for e.g. legacy projects.

\subsection{Keyphrase Extraction}
\label{sec:relatedkeyphrase}

Keyphrase Extraction (KE) is the task of locating a phrase or sentence that summarizes the content of an entire document~\cite{witten2005kea, zhao2011topical}.  KE is a subfield of extractive text summarization~\cite{gupta2010survey}. In text summarization generally, text of various lengths is extracted from longer documents, sometimes including multiple sentences from various locations of the document.  KE focuses exclusively on short phrases or sentences from (roughly) paragraph-sized documents.  Techniques for KE can be broadly categorized as either heuristic or neural-based.  Heuristic-based techniques include noun-phrase detection~\cite{barker2000using}, phrase ranking via network modeling~\cite{zhao2011topical}, manually-crafted features~\cite{nguyen2007keyphrase}, and rules based on statistical association among sentences~\cite{turney2003coherent}.  Recent surveys by Hasan~\emph{et al.}~\cite{hasan2014automatic} and Siddiqi~\emph{et al.}~\cite{siddiqi2015keyword} provide excellent coverage of these approaches.

As in many research areas, heuristic techniques have recently given way to neural-based techniques~\cite{zhang2016keyphrase, meng2017deep, basaldella2018bidirectional, subramanian2018neural, song2019abstractive, villmow2018automatic}.  While it has long been observed that KE techniques vary considerably across different application domains~\cite{frank1999domain}, these techniques generally follow a similar pattern: model a document in a word embedding space, highlight key phrases as training data, and train a neural network (usually a recurrent or convolutional net) with these highlighted phrases.  The idea is that the network will learn to recognize patterns of words that tend to indicate the start and end of keyphrases.  There is no clear ``best'' approach described in the literature; instead, different approaches have been shown to work in different domains e.g. Twitter comments~\cite{zhang2016keyphrase} versus academic articles~\cite{nguyen2007keyphrase}.


In this paper, rather than recreate a single existing extraction technique, we synthesize components that seem broadly effective in several domains -- thus, our work is related to all of these techniques but does not directly ``compete'' with any single alternative approach.  For example, Marujo~\emph{et al.}~\cite{marujo2015automatic} and Zhang~\emph{et al.}~\cite{zhang2016keyphrase} crowdsource the annotation of Twitter data via Amazon Turk in order to obtain training data for an RNN-based solution.  At a high level, this is a similar strategy to what we employ, except that we have an additional step of annotation from programming professionals, to ensure that the data collected by non-experts from Amazon Turk are consistent with annotations from experts.


\subsection{Crowdsourcing in Software Engineering}

Crowdsourcing is not a new concept in software engineering, though the specific application areas are quite diverse.  Stol~\emph{et al.}~\cite{stol2014two} provide an excellent look into how crowdsourcing is possible for software engineering tasks.  Their situation is quite different than what we propose: as Stol~\emph{et al.} point out, Amazon Turk typically involves breaking large tasks into \emph{micro-tasks} which tend to be short, repetitive, and require little domain knowledge.  In contrast, Stol~\emph{et al.} hire programmers via TopCoder to build specific pieces of a larger program.  Likewise, Yan~\emph{et al.}~\cite{yan2014itest} hire programmers in a crowdsourced process to improve software testing.  LaToza and van der Hoek~\cite{latoza2016crowdsourcing} discuss several similar strategies, involving hiring many programmers to assist in tasks that are relatively small in the context of software development.


The crowdsourcing problem in this paper is much more similar to the problem described in NLP literature for e.g. keyphrase annotation in Twitter data than to the crowdsourcing tasks typically described in SE literature.  Our task involves annotating a keyphrase in comments, which is often a simple matter of filtering formatting but also occasionally involves some degree of reading comprehension, even if it is not necessary to understand exactly what is being said.  For example, a non-expert can deduce that the phrase ``essentially what this function does is'' is a prelude to the summary, even if the non-expert does not understand the actual behavior described in the summary.  We echo the optimistic sentiment of Buhrmester~\emph{et al.}~\cite{buhrmester2011amazon} that suggests that properly-curated annotations from crowdsourced microtasks are a source of good-quality, inexpensive training data.

\subsection{Encoder-Decoder Architecture}
\label{sec:relatedencdec}

Our automated approach is based on an encoder-decoder model with attention used for Neural Machine Translation (NMT).  This architecture has been used extensively for a variety of tasks, and is thoroughly covered in related work~\cite{luong2015effective, johnson2017google, vaswani2017attention, lecun2015deep, bahdanau2014neural}.  In essence, an encoder-decoder NMT model is trained using pairs of sequences: an input sequence of word tokens in some source language, and an output sequence of word tokens in some target language. The purpose of the ``encoder'' is to generate a vector representation of the input sequence, which the ``decoder'' may then use to infer an output sequence one token at a time.  The key difference between regular NMT models and our model is that our model's output only needs to identify which words in the input sequence belong in the summary annotation  (see Section~\ref{sec:details} for details).

\begin{table*}[t!]
	\begin{centering}
		\caption{Annotations collected from all sources, pre- and post-filtering (see Section~\ref{sec:filter}).}
		\label{tab:annotations}
		\vspace{-0.1cm}
		\begin{tabular}{l|l|l|l|l}
			Set 				& 	Annotators 	& 	\# Functions 			& 	Annotations per Function 	&	\# Functions  after filtering 		\\ \hline
			&       			&     						&  						&        						\vspace{-0.2cm} \\
			Gold Set            	& 	Experts    		& 	1,000                   		& 	5          				& 980                 				\\
			Controlled Set     	& 	Non-Experts    	& 	20,000 (+ 1000 Gold)	& 	5    					& 18891 (+ 945 Gold) 			\\
			Expanded Set      	& 	Non-Experts   	& 	100,000  (+ 1000 Gold) 	& 	1 (5 for Gold)			& 97937(+ 980 Gold)  			\\
		\end{tabular}
		\vspace{-0.2cm}
		
	\end{centering}
\end{table*}

\section{Annotations}
\label{sec:annotation}

We performed annotation in three groups in this paper.  First, we created a \textbf{gold set} of 1000 C/C++ function summaries.  Our goal in creating the gold set is to provide a small, high-quality set of function comments and summaries against which to evaluate other approaches to extracting summaries.  Second, we created a \textbf{controlled non-expert set} of 20,000 summaries that were each annotated by five non-experts.  Third, we created an \textbf{expanded non-expert set} of 100,000 summaries that were annotated by one non-expert each.

\subsection{Corpus Preparation}

We prepared a corpus of C/C++ functions based on data provided by LeClair~\emph{et al.}~\cite{leclair2018adapting}.  LeClair~\emph{et al.} curate a dataset of around 25,000 C/C++ projects by applying standard pre-processing techniques such as removal of empty functions and identifier splitting.  To avoid errors and maintain consistency with related work, we use the preprocessed functions from their data (around 0.7m functions).  We randomly selected 1000 functions for the gold set, 20k for the controlled non-expert set, and 100k for the expanded non-expert set.  In keeping with recommendations for datasets for code summarization~\cite{leclair2018funcomdata}, we ensure that there is no overlap of functions from the same projects across sets to avoid contaminating the test set with information from the training set.  In other words, if a function \emph{A} is from project \emph{P}, and \emph{A} is in the gold set, then no functions from project \emph{P} will be in either of the non-expert sets.

\begin{figure}[b!]
	\vspace{-0.4cm}
	\includegraphics[width=8.8cm]{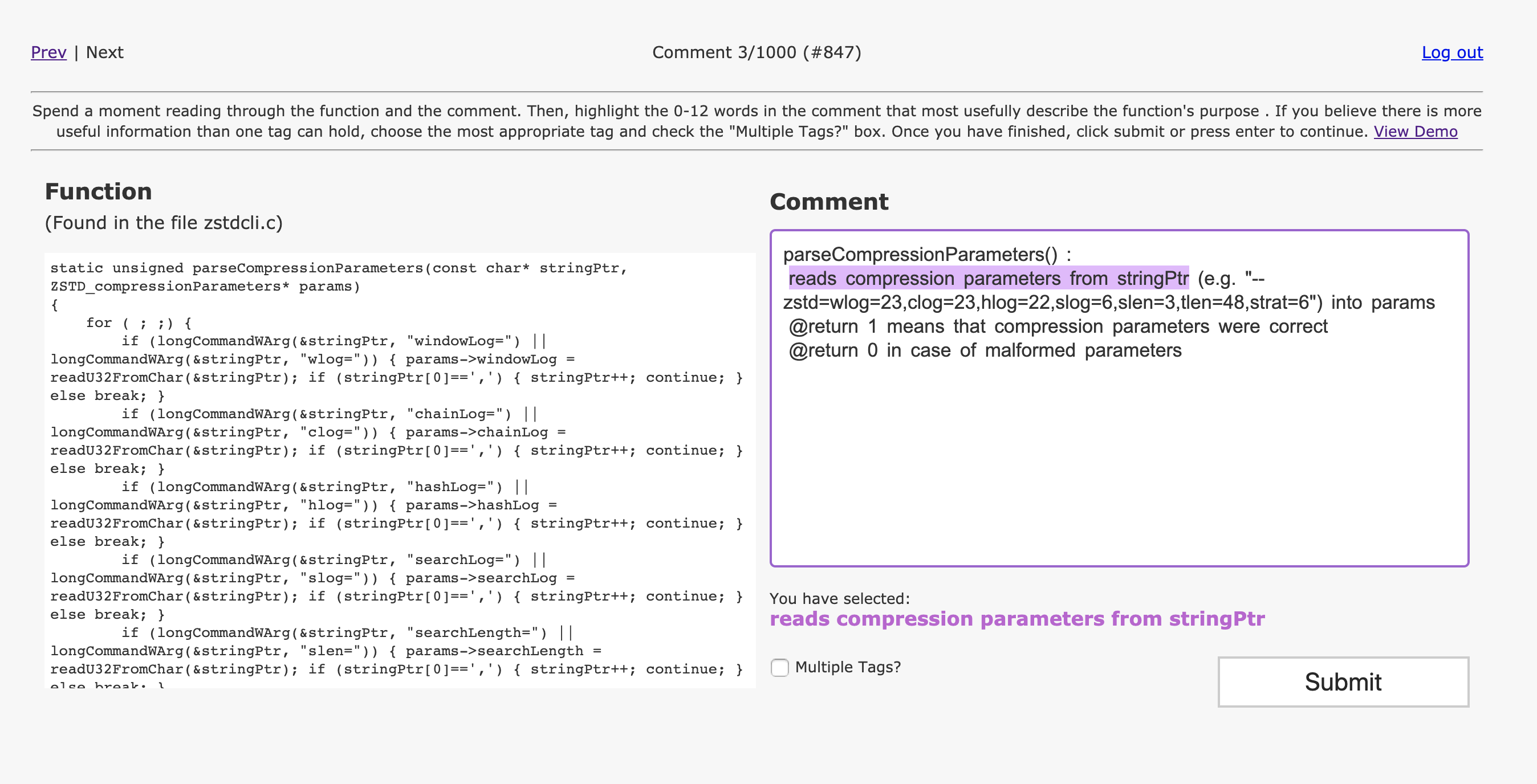}
	\vspace{-0.6cm}
	\caption{Screen capture of the annotation interface.  Non-expert annotators completed 21 function summaries on each page, highlighting the summary in the comment on the right.  Experts annotated one function per page.  The summary could be between zero and twelve words.}
	\label{fig:interface}
\end{figure}

\subsection{Annotation Interface}

We designed an interface to facilitate annotation.  A screen capture of the interface is shown in \Cref{fig:interface}.  For each function/comment pair, we gave annotators an interface displaying a C function on the left half of the screen and its corresponding comment on the right. Annotators were asked to highlight a contiguous span – up to 12 tokens – within the comment that best summarized the function. If a comment was found to contain no useful summary, the annotators were instructed not to highlight anything. We provided written instructions and several examples of comment/summary pairs in an instructions page that they could reference at any time. The functions appeared exactly as written, while the comments were formatted for readability. Specifically, the `/*' and `*/' strings at the start and end of the comment and extraneous `*' characters were removed, and we condensed contiguous spans of whitespace (spaces and tabs) to a single space (new lines were preserved). As annotators highlighted text, the words highlighted would appear in a different color beneath the comment to confirm the selection. If a function required more than 12 tokens to summarize, or multiple non-contiguous spans, annotators were instructed to highlight the most appropriate span and tick a checkbox reading ``Multiple Tags?''\footnote{In practice, we found that this checkbox was seldom marked. Therefore, we don't discuss the "Multiple tags?" data in the remainder of our analyses, but we do make the data available in our online appendix (Section~\ref{sec:reproducibility}).}. Once finished, annotators could click a Submit button to continue to the next tag(s).

\subsection{Gold Set Annotation}
\label{sec:annotategold}


We hired five professional programmers via Upwork~\cite{UPWORK} to each annotate all 1000 functions in the gold set.  The five programmers worked independently using the interface described in the previous subsection, and did not know that others would be annotating the same functions.  The result was 5000 unique annotations (five per function). The average time required was 5.8 hours per programmer to annotate all functions.  We compensated the programmers around US\$30 per hour.  Note that at this rate, it is prohibitively expensive to annotate a large repository of C/C++ functions -- it would cost around 29k hours of programmers time and US\$870k to annotate 1m C/C++ functions. 

\begin{figure}[b!]
	\vspace{-0.4cm}
	\includegraphics[width=8.8cm]{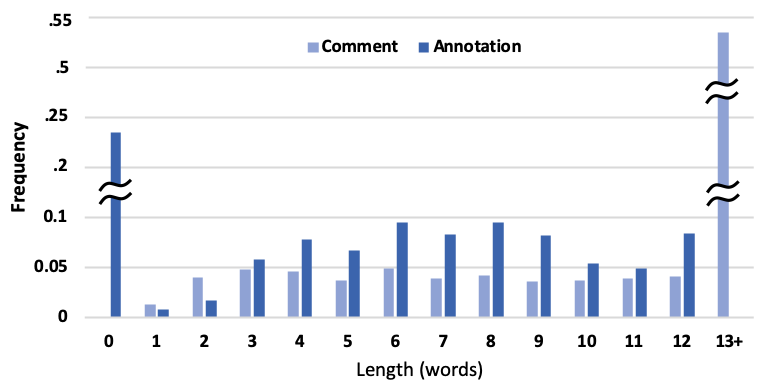}
	\vspace{-0.6cm}
	\caption{Length distribution of comments and annotations in the Gold set.}
	\label{fig:lengthdist}
\end{figure}

The length distribution of comments and expert annotations in the Gold set are shown in \Cref{fig:lengthdist}. There are two key observations: first, more than half of the comments are longer than 12 words. This means that a naive summary extraction approach that only takes the first 12 words would blindly exclude content from more than half of these comments. Second, over 20\% of the expert's annotations indicated that there was no valid summary in a comment, meaning that those comments may not be good candidates for downstream tasks.

\subsection{Controlled Non-Expert Annotation}
\label{sec:annotatecne}

We used Amazon Mechanical Turk~\cite{MTURK} to recruit non-expert workers to perform the ``microtask'' (also known as a HIT) of reading and annotating a function comment using the annotation interface described above.  We paid five cents per microtask for each of the 20k functions in the set.  We call this set of annotations ``controlled'' because we obtain five unique annotations per function -- this redundancy permits us to answer research questions about the consistency of the annotations (see Section~\ref{sec:rqs}) and to ``vote'' together annotations to create a single annotation of higher quality (see Section~\ref{sec:voting}). We also obtained annotations for the 1k gold set functions, in order to compare the consistency of non-experts to the professional annotations.  

The non-expert workers were not required to have any prior programming knowledge; as mentioned in \Cref{sec:intro}, a key intuition is that summary extraction generally does not require interpretation of programming concepts, only filtering of obviously unrelated material. They were informed of the nature of the task prior to attempting it, and were provided written instructions and examples. To filter out potential spammers, participants were required to have completed at least 50 prior tasks on AMT with at least a 95\% approval rating, as recommended by Peer~\emph{et al.}~\cite{Peer2014}. 

For the non-experts, the annotation interface showed 21 functions per page instead of one. This increased number of functions per page was due to restrictions of the Amazon Turk system combined with the need to include \textbf{Quality Control (QC)} functions.  We include one function from the gold set as a QC function among the 20 other functions. Note that at the rate of 25 cents per function (five cents per annotation, times five annotators), the cost to annotate 1m functions would be US\$250k: quite a lot less than professionals, but an amount that may still be prohibitively expensive.  


\subsection{Expanded Non-Expert Annotation}

We used the same process to build the expanded non-expert annotation set as in the controlled non-expert set.  However, we added a further constraint that the annotators could not be given the same QC function twice.  Since we sought 100k annotations and only have 1k gold set annotations, it is not possible to give a unique QC function for every 20 functions.  To avoid a situation in which an annotator saw a QC function more than one time, we obtained the annotations in five groups of 20k functions.  After each group, we prohibited any annotator from previous groups from participating again. Due to a technical error, the first group was only given a single QC function, repeated in every set of 21 functions.  Therefore, annotators in the first group were allowed to participate in one additional round of annotation. At a cost of five cents per microtask, 100k annotations costs US\$5000. One million annotations would cost US\$50,000, approaching a level at which it may be affordable to annotate an entire large dataset.

\subsection{Dataset Filtering}
\label{sec:filter}

Some problems occurred during annotation that necessitated filtering some results from the datasets.  These problems are unfortunately typical of work involving humans via Amazon Mechanical Turk and crowdsourcing systems in general.  Specifically, we found that a small number of comments from the gold set and non-expert sets included non-comment text, such as ``void /* PRIVATE */'' and ``putchar(' '); /*space out the next line */'', or contained special characters, causing the comments to appear incorrectly for the participants.  We chose to filter out these problematic comments and their annotations, leaving the ``1k'' gold set with 980 functions, the ``20k'' set with 18811, and the expanded set with 97937. 

Upon inspection, we discovered that one of the five expert programmers produced a significant number of annotations that were plainly incorrect (e.g. an annotation consisting only of the word ``so,'' where the comment clearly contains a succinct summary). Although the majority of this expert's results were in agreement with the others, the presence of multiple egregious annotations was a cause for concern. To maintain the quality of the gold set, we chose not to include any of this annotator's annotations in our subsequent analyses, though they are available in our online appendix (Section~\ref{sec:reproducibility}). 

Furthermore, we took a number of measures to filter out spam and other unsuitable results from Mechanical Turk. The aim here was not to remove all annotations that we deemed to be ``low-quality,'' as we wanted our subjective interpretations to bias the non-expert set as little as possible. Rather, the aim was to filter out workers who failed to follow the basic annotation protocols, either by 1) frequently providing plainly incorrect annotations (e.g. consisting only of the word ``the'') or 2) frequently annotating summaries in comments where it was clear that no summary existed (e.g. in comments consisting only of the function signature). Therefore, after collecting the results from Mechanical Turk, we ran a script to automatically flag any worker exhibiting one of the following behaviors:

\begin{itemize}
  \item Averaging less than 50\% similarity with the experts on the QC questions.
  \item Failing to mark the ``No tag?'' checkbox at least once.
  \item Completing submissions in $< 120$ seconds on average. 
\end{itemize}

The first author manually reviewed the flagged workers and rejected all submissions from any worker who failed to follow the basic annotation procedures as described above. 

\subsection{Agreement / Voting Procedure}
\label{sec:voting}

We applied an agreement voting procedure in order to create a set of \textbf{unified expert annotations} and a set of \textbf{unified controlled non-expert annotations}.  The goal is to create a single ``best'' annotation for each situation in which we have multiple annotations (e.g. four expert annotations for each of the 1k functions in the expert-annotated gold set).  



There is no single accepted procedure for precisely our situation, though sophisticated strategies exist for related problems; Zheng~\emph{et al.}~\cite{zheng2017truth} survey more than a dozen methods addressing the ``truth inference'' problem in crowdsourcing. One straightforward method would be to take a majority vote on a word-by-word basis, reconstructing a unified summary from each token that appeared in the majority of the annotations. But a problem with that application is that it could result in annotations that none of the annotators actually selected -- an ``average'' annotation that doesn't represent a valid selection.  

Instead, our approach works by choosing one annotation among the set of annotations for each function.  Technically, we calculate a probability $P(A|W)$ for each word $w$ in a comment either occurring in the annotation (event $A$) or not.  Then for each annotation (span $s$ in a comment selected by an annotator) for a function, compute $P(S=s) = \prod_{i=0}^{n} P(A=A_{i} | W=i)$.  Finally, we select the span $s$ with the highest probability as the span for the unified set. It was possible for multiple spans to ``tie'' with equal probability. Ties among expert annotations in the gold set were decided by the first author, and ties among non-experts were resolved by randomly choosing one of the tied options, so as not to impart a bias towards longer or shorter annotations.

\section{Automated Annotation Prediction}
\label{sec:auto}

We use three automated techniques to predict annotations given a function comment.  All three are of our design, but are based on a synthesis of related work.  Since, to our knowledge, no directly competitive solution exists for annotating summaries in unstructured function comments, we design two approaches based on solutions to related problems, in addition to one baseline heuristic.

\subsection{Baseline: First Twelve Tokens}

A naive baseline is to use the first \emph{n} tokens from the comment (\emph{n}=12 in this paper, see end of Section~\ref{sec:intro}).  While we do not expect it to perform well in terms of precision, it serves as an important comparison due to its simplicity as a heuristic.  Plus, it is likely to achieve high recall, since a majority of summaries occur near the start of the comment and are less than twelve words long.



\subsection{BiLSTM: Summary from Comment Only}


This approach utilizes a bidirectional long short-term memory (BiLSTM) architecture. It represents a synthesis of state-of-the-art solutions from the natural language processing (NLP) research community.  As Section~\ref{sec:relatedkeyphrase} discussed, keyphrase extraction is an active research area.  The problem is defined as selecting a sentence summary out of a longer document.   In general, the state-of-the-art techniques involve a neural model in which the document and the summaries are inputs used to train an encoder-decoder architecture based on recurrent networks (see Section~\ref{sec:relatedencdec} for a high-level discussion of this architecture).  Nearly all recent techniques are based on a bidirectional RNN in the encoder and a unidirectional RNN in the decoder, though several smaller design decisions make it implausible to create one ``best'' approach for every situation.  Therefore, we built an approach that synthesizes several design decisions that appear to be broadly effective for different problems.  We provide details and justification in the implementation section below.

\subsection{BiLSTM+F: Summary from Comment and Source Code}

Our third approach is based on a mixture of NLP and SE related work that considers a representation of the source code in addition to the comment.  The BiLSTM approach uses only the comment in order to make a decision about the keyphrase for that comment. However, the human annotators in Section~\ref{sec:annotation} saw both a function's comment and its source code.  Even if the annotators do not understand the function, different words and phrases in the code could provide clues about the summary.  At the same time, several recent papers (e.g. \cite{iyer2016summarizing, leclair2018funcom, fernandes2018structured, alon2018code2seq}) have explored neural-based representations of source code for the task of summarization.  We integrate a recent representation described by LeClair~\emph{et al.}~\cite{leclair2018funcom} at ICSE'19 into the BiLSTM approach.  Essentially, we augmented the encoder to accept the function source code as another input alongside the comment, but otherwise left the BiLSTM the same.  We call the approach BiLSTM+F for BiLSTM plus Function code. 

\subsection{Model Details and Implementation}
\label{sec:details}

To promote clarity and reproducibility, we describe the BiLSTM and BiLSTM+F approaches in the context of the Keras source code that we wrote to implement them (slightly edited for readability).  Because the two approaches are so similar, we present the code for BiLSTM+F, and point out the lines which are not in BiLSTM.  All code and related infrastructure are available in our online appendix (Section~\ref{sec:reproducibility}).

We train the encoder-decoder model using comment/function sequences as input (just comments for BiLSTM) and annotations as the output.  The annotation output is a sequence of the same length as the comment.  At each position is one of five values for the word in the comment at the same position: 1) the word is in the annotation, 2) it is not in the annotation, 3) start sequence token, 4) end sequence token, or 5) padding.  E.g., for an input sequence [$<$s$>$, a, b, c, d, $<$/s$>$] a sample output vector might be [3, 2, 1, 1, 1, 4] if the annotation were ``b c d.''

The encoder consists of two parts: the comment encoder and the function encoder.  The comment encoder accepts the entire comment as input and is in both BiLSTM and BiLSTM+F.  Note we use a word embedding space of 100 units length and vocabulary size of 2000.  We established these values based on related work~\cite{leclair2018funcom}, and by conducting exploratory pilot studies.  This part is a traditional RNN encoder:

\vspace{-0.1cm}
{\small \begin{verbatim}
enc_com_in = Input(shape=(100,))
enc_com_emb = Embedding(output_dim=100, 
    input_dim=2000)(enc_com_in)
enc_com = Bidirectional(CuDNNLSTM(100,
    return_state=True, return_sequences=True))
enc_com_out, f_h, f_c, b_h, b_c =
    encoder_com(enc_com_emb)
com_state_h = concatenate([f_h, b_h])
com_state_c = concatenate([f_c, b_c])
enc_com_states = [com_state_h, com_state_c]
	\end{verbatim}}
\vspace{-0.1cm}

In BiLSTM+F, we also encode the function source code.  As recommended in related work~\cite{leclair2018funcom}, we use an input length of 100 tokens, though for space constraints we limit the vocabulary size to 2000 from 10k (we did not notice significant performance degradation for this problem domain).  Two other differences from LeClair~\emph{et al.}'s encoder are that 1) we use a bidirectional RNN instead of unidirectional, and 2) we use an LSTM instead of a GRU.  Again, we chose this architecture by considering the results of our pilot studies, though we did not observe major performance differences in the end.

\vspace{-0.1cm}
{\small \begin{verbatim}
enc_src_in = Input(shape=(100,))
enc_src_emb = Embedding(output_dim=100, 
    input_dim=2000)(enc_src_in)
enc_src = Bidirectional(CuDNNLSTM(100,
    return_state=True, return_sequences=True))
enc_src_out, f_h, f_c, b_h, b_c =
    encoder_src(enc_src_emb)
src_state_h = concatenate([f_h, b_h])
src_state_c = concatenate([f_c, b_c])
enc_src_states = [src_state_h, src_state_c]
\end{verbatim}}
\vspace{-0.1cm}

Next, we use a unidirectional decoder to represent the annotation.  Note that the length of the RNN output vector is double the encoder's, to accommodate both directions of the encoder.  Also note that the vocabulary size is only 5.  An important detail to our approach is that we do \emph{not} attempt to predict every word of the annotation, despite the overall similarity of our approach to a seq2seq NMT approach.  Instead, we predict one of five values for each word, as detailed earlier.  The advantage is that the smaller vocabulary size makes the prediction much easier for the network to learn.

\vspace{-0.1cm}
{\small \begin{verbatim}
dec_in = Input(shape=(12,))
dec_emb = Embedding(output_dim=100,
    input_dim=5)(dec_in)
decoder = CuDNNLSTM(200, 
    return_state=True, return_sequences=True)
dec_out, _, _ = decoder(dec_emb,
    initial_state=enc_com_states)
\end{verbatim}}
\vspace{-0.1cm}

Next, we implement the attention mechanism described by Luong~\emph{et al.}~\cite{luong2015effective} to attend output predictions to words in the encoder.  Since we are predicting whether comment tokens appear in an annotation (and not generating text, as in the usual NMT use of attention), we do not create a ``dictionary'' of output to input words.  Instead, the attention mechanism should help identify words likely to be in the annotation.

\vspace{-0.1cm}
{\small \begin{verbatim}
com_attn =
    dot([dec_out, enc_com_out], axes=[2, 2])
com_attn = Activation('softmax')(com_attn)
com_context =
    dot([com_attn, enc_com_out], axes=[2,1])
\end{verbatim}}
\vspace{-0.1cm}

We also implement attention over the source code encoding, though the result of this attention is admittedly harder to interpret.  In general, we expect word occurrence in the source code to help identify whether a word is in an annotation.

\vspace{-0.1cm}
{\small \begin{verbatim}
src_attn =
    dot([dec_out, enc_src_out], axes=[2,2])
src_attn = Activation('softmax')(src_attn)
src_context =
    dot([src_attn, enc_src_out], axes=[2,1])
\end{verbatim}}
\vspace{-0.1cm}

Finally, we concatenate the context matrices and create a fully-connected output layer following standard procedure for seq2seq encoder-decoder models.    

\vspace{-0.1cm}
{\small \begin{verbatim}
context = concatenate(
    [com_context, dec_output, src_context])
context = TimeDistributed(
    Dense(300, activation="relu"))(context)
dec_dense = Dense(5, activation="softmax")
dec_out = dec_dense(context)
train_model = Model(
    inputs=[enc_com_in, enc_src_in, dec_input],
    outputs=dec_output)
encoder_model = Model(
    inputs=[enc_com_in, enc_src_in],
    outputs=[enc_com_out, enc_com_states, 
               enc_src_out])
\end{verbatim}}
\vspace{-0.1cm}

Note again that the outputs are not words in a vocabulary; they are predictions of whether each word in the input comment is or is not in the summary.

\section{Experiments}
\label{sec:experiments}

This section discusses our experiments, including research questions, methodology, and other key settings.

\subsection{Research Questions}
\label{sec:rqs}

Our research objective is to evaluate inexpensive non-expert human annotation and automated machine annotation as alternatives to expensive annotation by expert humans.  Towards this end, we ask the following Research Questions:

\begin{description}
	\item[$RQ_1$] How similar are the annotations by individual experts to the unified set of expert annotations?
	\vspace{0.1cm}
	\item[$RQ_2$] How similar are the annotations by individual non-experts to the unified set of non-expert annotations?
	\vspace{-0.325cm}
	\item[$RQ_3$] How similar are the annotations in the unified controlled non-expert set to the unified expert set?
	\vspace{0.1cm}
	\item[$RQ_4$] How similar are the annotations in the expanded non-expert set to the unified controlled non-expert set?
	\vspace{0.1cm}
	\item[$RQ_5$] How similar are the annotations in the expanded non-expert set to the unified expert set?
	\vspace{0.1cm}
	\item[$RQ_6$] How similar are the annotations by the automated techniques to the unified controlled non-expert set?
	\vspace{0.1cm}
	\item[$RQ_7$] How similar are the annotations by the automated techniques to the unified expert set?
\end{description}

RQ$_1$, RQ$_3$, RQ$_5$, and RQ$_7$ compare annotations to the unified expert set, while RQ$_2$, RQ$_4$, and RQ$_6$ compare annotations to the unified controlled non-expert set. All comparisons are made using annotations for the set of 945 gold set functions that were annotated by experts and non-experts and that remained after the filtering process described in \Cref{tab:annotations}. 

Note that in our discussion of these RQs we use the term ``similar'' quite generally.  In principle, we consider annotations to be better when they are more similar to the unified set of expert annotations. See Section~\ref{sec:metrics} for the specific metrics we use to measure similarity.

The rationale for $RQ_1$ is to provide a baseline for understanding the similarity of non-experts and machine annotations to the experts' annotations.  We do not expect the experts to agree on every annotation.  While we use a voting mechanism (Section~\ref{sec:voting}) to produce a unified set of expert annotations, it is difficult to interpret the similarity of non-expert and automated annotations to this unified set, without knowing the similarity of the experts to the unified set. Likewise, the rationale for $RQ_2$ is to provide a baseline for understanding the quality of the machine annotations. The rationale for $RQ_3$, $RQ_4$, and $RQ_5$ is to evaluate the quality of the two non-expert human-based alternatives.  We expect the unified controlled set to be of higher quality than the expanded set, but it is possible that they are quite similar.  If they are, then the expanded procedure provides a much more cost-effective human-based annotation than the controlled set. Finally, we ask $RQ_6$ and $RQ_7$ to evaluate the quality of the automated techniques.  The automated techniques provide an enormous opportunity to reduce costs, but at an unknown quality penalty.


 
 

\subsection{Methodology}

Our methodology for answering our research questions is straightforward: we calculate similarity metrics specified in the next section for different groups of annotations.  For $RQ_1$ and $RQ_2$, we report averages of metrics for each annotator to the relevant unified set. We also report outliers and present illustrative examples.  For $RQ_3$, we directly compare the unified controlled non-expert set to the unified expert set. For $RQ_4$, and $RQ_5$, we compare the expanded non-expert set to the unified non-expert set and the unified expert set, respectively. For $RQ_6$ and $RQ_7$, we compute similarity metrics of each automated approach to the unified sets.


\subsection{Metrics}
\label{sec:metrics}

To the best of our knowledge, there is no single metric that fully captures the notion of ``similarity'' in the context of this constrained annotation task. While it can be described as extractive summarization, it differs from similar tasks in that annotators can elect not to provide a summary. Additionally, annotators are limited to selecting a single contiguous span of text of up to 12 tokens, whereas other extractive summarization tasks generally allow summaries to be built from multiple, discontiguous spans. Therefore, we use three separate metrics to capture different aspects of similarity. Specifically, we measure similarity by framing our task as an \textit{extractive summarization} task (measured by ROUGE), a \textit{binary classification} task (measured by recall, precision, and F1), and a \textit{unitization} task (measured by Krippendorff’s unitized alpha). We report mean scores for all annotations in the test sets.


\subsubsection{Extractive Summarization}
We use two configurations of the ROUGE metric~\cite{lin2004rouge} to directly measure the similarity between two summaries -- ROUGE-1 and ROUGE-L.  ROUGE is a long-accepted metric for evaluating sentence-length summaries of documents. Lin~\emph{et al.}~\cite{lin2003automatic} have demonstrated that summary evaluation based on simple unigram overlap (ROUGE-1) correlates highly with human judgment. 

In addition to reporting ROUGE scores as given in related work, we report F1 measure for ROUGE-1-NE and ROUGE-L-NE, which we define as the ROUGE scores for the predictions which are not empty (NE).  There are cases in which an annotator or automated approach predicts no summary when in fact a summary exists.  These cases reduce the ROUGE scores, but leave open a question about how good the annotations are when a prediction \emph{is} made.  A desirable characteristic of an automated approach is to make no prediction instead of making a poor prediction, so that predictions from the approach will be trusted.  Therefore, we report F1 scores for these cases.  We do not report precision and recall for these due to space limitations since they do not affect our conclusions, but they are available in our online appendix (Section~\ref{sec:reproducibility}).

\subsubsection{Binary Classification}
ROUGE metrics can only be calculated when there is a valid reference string. While they can compare an empty hypothesis to a non-empty reference (resulting in a score of 0), they cannot compare a hypothesis to an empty reference. Therefore, we separately evaluate the binary classification task of identifying comments that do not contain valid summaries. To measure classification similarity between a hypothesis set and a reference set, we report the conventional precision, recall, and F1 measures (treating comments without valid summaries as the positive class).

\subsubsection{Unitization}
The task of segmenting continuous data into mutually exclusive sections, each belonging to a positive class (a ``unit'') or negative class (a ``gap''), is known as unitization. Comparing ``summary labels'' as ``comment unitzations'' allows us to incorporate aspects of both previous metrics, as well as additional information about the ``difficulty" of choosing a correct annotation (i.e. annotators are less likely to choose the same label given a long comment than given a short comment).


To measure the similarity of different unitizations of the same data, we use Krippendorff's alpha, $\alpha$, a generalized, chance-corrected measure of inter-annotator agreement \cite{krippendorff1995reliability}. Unlike other agreement metrics that can only measure reliability in the coding of predetermined units, Krippendorff's alpha has an adaptation called the ``unitized alpha'' $\alpha_U$ which is widely used to measure reliability in the unitizations themselves \cite{krippendorff2004measuring, ghosh2014analyzing, stab2014annotating}. Metrics from the $\alpha$ family are calculated by subtracting the chance-corrected level of disagreement (that is, the ratio of the disagreement observed to the disagreement expected) from $1$. An $\alpha$ score of $1$ indicates that annotators are in perfect agreement (i.e. they have chosen identical annotations); a score of 0 indicates a level of agreement expected by random chance (i.e. there is likely some, but not much overlap between the annotations); and a score less than 0 indicates worse agreement than is expected by chance (i.e. there is little, if any, overlap).



\subsection{Threats to Validity}

Like all experiments, our work carries threats to validity.  Key threats include: 1) The effect of the voting procedure.  We attempt to mitigate this risk by using accepted agreement metrics and by studying the similarity of individual annotators to the unified sets ($RQ_1$ and $RQ_2$), but a risk remains that a different voting procedure would cause different results.  2) Human factors.  Any work by human annotators is subject to human factors such as fatigue, previous experience, and bias.  We attempt to mitigate this with a voting procedure in the expert and controlled non-expert sets, but a threat still exists that different participants would yield different results.  3) Source code.  We use a large repository of source code, but it is possible that different code would yield different results.

\begin{table*}[]
	\begin{centering}
	\caption{Statistical summary for RQ$_1$: similarity of expert annotators to unified expert set.}
	\label{tab:rq1}
	\vspace{-0.1cm}
	\begin{tabular}{p{2.5cm}|llll|llll|lll|lll|lll}
		\multicolumn{1}{c}{} & \multicolumn{4}{c}{ROUGE-1} & \multicolumn{4}{c}{ROUGE-L} & \multicolumn{3}{c}{is-empty} & \multicolumn{3}{c}{unitization} \\
		& P  & R  & F1  & F1-NE & P  & R  & F1 & F1-NE  & P  & R  & F1  &  $\alpha_u$   \\ \hline
				       &   &     &     &  &        &   &  &  &     &     & &      \vspace{-0.2cm} \\
		Max. & 93.24 & 95.14 & 93.48 & 96.47 & 93.64 & 95.30 & 93.93 &96.85& 90.64 & 97.13 &         89.32 &  0.77 \\
		Median & 91.98 & 91.52 & 91.20&95.78 & 92.21 & 91.83 & 91.61 & 96.24 & 83.82 & 91.39 &         87.39 &  0.67 \\
		Min. & 88.64 & 86.91 & 87.22 & 95.24 & 88.79 & 87.28 & 87.61 & 95.67 & 76.60 & 88.04 &         85.65 &  0.63 \\
		Mean & 91.46 & 91.27 & 90.77 & 95.81 & 91.71 & 91.56 & 91.19 & 96.25 & 83.72 & 91.99 &         87.44 &  0.69 \\
		Stddev &  1.98 &  3.37 &  2.60 &  0.52 & 2.08 &  3.29 &  2.63 &  0.51 & 5.79 &  4.66 &          1.71 &  0.06 \\
		\end{tabular}

	\vspace{0.3cm}
	\caption{Similarity values for RQ$_2$ - RQ$_5$.}
	\label{tab:rq25}
	\vspace{-0.3cm}

	\begin{tabular}{p{2.5cm}|llll|llll|lll|lll|lll}
		\multicolumn{1}{c}{} & \multicolumn{4}{c}{ROUGE-1} & \multicolumn{4}{c}{ROUGE-L} & \multicolumn{3}{c}{is-empty} & \multicolumn{3}{c}{unitization} \\
		& P  & R  & F1  & F1-NE & P  & R  & F1 & F1-NE  & P  & R  & F1  &  $\alpha_u$   \\ \hline
				       &   &     &     &  &        &   &  &  &     &     & &      \vspace{-0.2cm} \\
		RQ$_2$ (mean) & 86.92 & 86.63 & 86.14 & 92.40 & 87.22 & 86.97 & 86.62 & 92.92 & 75.65 & 75.56 &         73.75 &  0.43 \\
		RQ$_3$ & 92.19 & 88.09 & 89.07 & 92.59 & 92.41 & 88.79 & 89.79 & 93.34 & 85.49 & 78.95 &         82.09 &  0.66 \\
		RQ$_4$ & 84.86 & 80.72 & 81.52 & 86.70 & 85.17 & 81.51 & 82.38 &  87.82 & 70.65 & 65.67 &         68.07 &  0.32 \\
		RQ$_5$ & 81.22 & 80.99 & 79.62 & 87.87  & 81.85 & 81.65 & 80.62 &  88.82 & 66.79 & 67.23 &         67.01 &  0.35 \\

	\end{tabular}

	\vspace{0.3cm}
	\caption{Similarity values for RQ$_6$: automated approaches to non-expert test set.}
	\label{tab:rq6}
	\vspace{-0.3cm}

	\begin{tabular}{p{2.5cm}|llll|llll|lll|lll|lll}
		\multicolumn{1}{c}{} & \multicolumn{4}{c}{ROUGE-1} & \multicolumn{4}{c}{ROUGE-L} & \multicolumn{3}{c}{is-empty} & \multicolumn{3}{c}{unitization} \\
		& P  & R  & F1  & F1-NE & P  & R  & F1 & F1-NE  & P  & R  & F1  &  $\alpha_u$   \\ \hline
				       &   &     &     &  &        &   &  &  &     &     & &      \vspace{-0.2cm} \\
		Baseline: 12 Tokens       & 67.79       & 94.27    & 76.46 & N/A & 70.89       & 94.05    & 79.03 & N/A & N/A       & N/A    & N/A  & --0.43  \\
		BiLSTM  & 72.68 & 78.09 & 73.94 & 86.60 & 73.91 & 78.62 & 75.18 & 88.07 & 55.65 & 71.50 &         62.59 &  0.36 \\
		BiLSTM+F       & 67.12 & 71.02 & 67.63 & 86.64 & 68.15 & 71.60 & 68.79 & 88.12 & 45.54 & 71.50 &         55.65 &  0.19 
	\end{tabular}

	\vspace{0.3cm}
	\caption{Similarity values for RQ$_7$: automated approaches to expert test set (gold set).}
	\label{tab:rq7}
	\vspace{-0.3cm}

	\begin{tabular}{p{2.5cm}|llll|llll|lll|lll|lll}
		\multicolumn{1}{c}{} & \multicolumn{4}{c}{ROUGE-1} & \multicolumn{4}{c}{ROUGE-L} & \multicolumn{3}{c}{is-empty} & \multicolumn{3}{c}{unitization} \\
		& P  & R  & F1  & F1-NE & P  & R  & F1 & F1-NE  & P  & R  & F1  &  $\alpha_u$   \\ \hline
				       &   &     &     &  &        &   &  &  &     &     & &      \vspace{-0.2cm} \\
		Baseline: 12 Tokens       & 71.21       & 94.25    & 78.85 & N/A & 73.92       & 94.04    & 81.07 & N/A & N/A       & N/A    & N/A  & --0.52  \\
		BiLSTM & 76.60 & 78.99 & 76.81 & 88.89 & 77.49 & 79.55 & 77.79 &  90.03 &59.68 & 70.81 &         64.77 &  0.36 \\
		BiLSTM+F & 71.25 & 71.74 & 70.34 & 88.04 & 71.97 & 72.43 & 71.34 & 89.30 & 51.16 & 74.16 &         60.55 &  0.22 
	\end{tabular}
	
	\end{centering}
	\vspace{-0.2cm}
\end{table*}

\section{Experimental Results}
\label{sec:results}

We provide results of the experiment described in the last section to answer our research questions. 

\subsection{RQ$_1$: Similarity within Gold Set}

We found that the experts tended to have a high degree of similarity with the unified gold set.  We created the unified gold set by choosing the most probable annotation from any one expert (as described in ~\Cref{sec:annotategold}), so it is given that at least one annotator will ``agree'' completely with the gold set for every function.  ~\Cref{tab:rq1} summarizes the ROUGE scores for all of the five annotators.  There was one annotator responsible for a high proportion of selected annotations and by one measure could be considered the ``best'' annotator (93.48\% F1 ROUGE-1 score).  However, even the annotator with the lowest similarity with the unified set by the same measure (87.22\% F1 ROUGE-1) was not that different.  Manual inspection of the annotations reveals that many of these cases emerged from relatively minor disagreements e.g. whether to annotate ``this function processes...'' or ``processes...'' as the summary.  An important conclusion is that similarity measures for RQ$_1$ form a ceiling of expected performance for RQ$_3$ - RQ$_7$: for example, a ROUGE-1 F1 score in the low 90\% range would be very high for the non-experts and automated approaches, considering that the experts themselves are in the same range.

\subsection{RQ$_2$: Similarity within Controlled Non-Expert Set}



\begin{figure}[b!]
	\vspace{-0.4cm}
	\includegraphics[width=8cm]{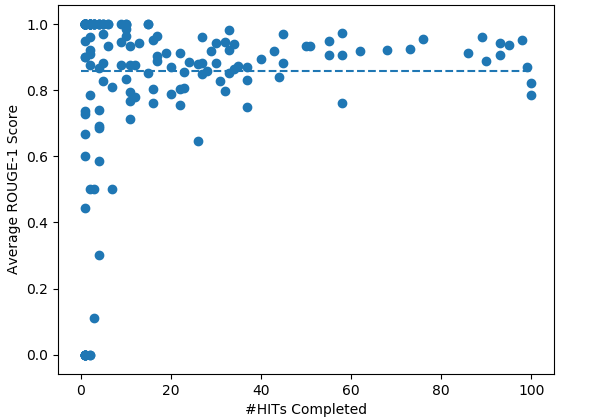}
	\vspace{-0.2cm}
	\caption{Performance of individual Mechanical Turk workers in RQ$_2$ as measured by ROUGE-1 F1, compared to the total number of HITs completed by each worker. The dashed line signifies the mean F1 score.}
	\label{fig:rq2}
\end{figure}

We found that the non-experts in the controlled non-expert set had a relatively high level of similarity to the unified controlled non-expert set.  In general, similarity is a bit lower (~5\%) than that observed for RQ$_1$ (see RQ$_2$ line in \Cref{tab:rq25} and Mean line in \Cref{tab:rq1}), with a notable difference that the non-experts disagreed much more often about whether any annotation exists at all (see is-empty column).  After manual inspection of the results, our working hypothesis is that annotators probably ``felt a need'' to annotate something in ambiguous cases, and lacked the expert knowledge necessary to be confident in leaving an annotation blank.  A somewhat humorous example is the comment for function id 1073101 in the gold set ``this is a hack, but I guess that's what I writing [\emph{sic}] anyways.''  The experts unanimously agreed that no summary exists in this comment (i.e. a blank annotation).  But the unified non-expert set annotation is ``this is a hack.''  The connotation of the word ``hack'' was a clear signal to the experts, but not the non-experts.

Consistent with other literature using Mechanical Turk~\cite{Grady:2010:CDR:1866696.1866723}, we observe that many workers who complete only a few HITs exhibit poor performance, while those who complete more HITs consistently perform reasonably well, as shown in \Cref{fig:rq2}. Given this variation, it could be misleading to report a statistical summary of all annotators as we did for RQ$_1$, so we summarize the results as a mean for general comparison on all metrics in \Cref{tab:rq25}. 

A takeaway is that for consistent results, researchers wishing to obtain consistent results from non-expert crowdsourced summary annotation may wish to require each worker to complete a minimum of 200 annotations (10 tasks, recall each task contained 20 function comments to annotate), as performance varies considerably below that threshold.

\subsection{RQ$_3$: Unified Controlled Non-Expert to Unified Gold Set}

The observed similarity between the unified controlled non-expert set and the unified gold set was just slightly below, but generally in line with, the mean similarity between the individual experts and the unified gold set.  Recall that we asked the non-experts to annotate one function from the gold set for every 20 that they annotated from the controlled set (\Cref{sec:annotatecne}), resulting in multiple annotations for each function, which were condensed into the unified controlled non-expert set\footnote{Due to the filtering performed on the controlled non-expert set, some functions were left only one or two annotations. We expect that comparing only the functions voted upon by several annotators would result in higher agreement with the experts, as demonstrated in related research~\cite{zheng2017truth}.}.  For these gold set functions, the unified controlled set performed nearly as well as any individual expert: 89.07\% ROUGE-1 F1 vs expert mean 90.77\%, 89.79\% ROUGE-L F1 vs 91.19\%, etc. (see RQ$_3$ line in \Cref{tab:rq25} and Mean line in \Cref{tab:rq1}).  The takeaway is that, in aggregate, a unified set of non-expert annotations is comparable to an expert annotator in our problem domain.

\subsection{RQ$_4$: Expanded Set to Unified Controlled Non-Expert Set}

The expanded set achieves slightly lower and more variable similarity to the unified controlled set than the average of annotators within the controlled set. Recall that, while the expanded set consists of one non-expert annotation for each of ~100,000 functions, the gold set QC functions are still labeled by multiple expanded set annotators, drawn from the same worker pool as the controlled non-expert set. We would, therefore, expect comparable performance of individual expanded set annotators to individual controlled set annotators on this set of functions. One factor contributing to the lower reported similarity is that the unified set was selected from the annotators in the controlled set, so at least one entry will have 100\% similarity, affecting the average. A key observation is that the variation is largely at the edges of the annotation, with disagreements like the ``this function'' example in RQ$_1$.  This factor is observed in comparing the ROUGE-1 scores to ROUGE-L: ROUGE-1, which measures individual words, is slightly lower than ROUGE-L, which measures common sequences. Small differences like these may cause the apparent inflated disagreement between the expanded set annotations and the unified controlled set.

\subsection{RQ$_5$: Expanded Set to Unified Gold Set}

A similar pattern emerges when comparing the expanded set to the gold set as to the unified controlled set: overall general agreement is relatively high (ROUGE-L scores compared to experts in RQ$_1$ are about 10\% lower) but variation increases at the edges of the annotations.  Whether this variation is tolerable depends on one's application: if only a rough annotation of functions in legacy code is needed for documentation, a procedure akin to the one we used for the expanded set may be sufficient, at a very considerable cost reduction compared to hiring experts (see \Cref{sec:annotation} for cost estimates).

\subsection{RQ$_6$ and RQ$_7$: Performance of Automated Approaches}

We found that the best automated approach was BiLSTM.  The BiLSTM+F model has slightly lower performance, perhaps due to the model including many more parameters, thus requiring more data to train: encoding of function sequences in other literature are trained on millions of examples, versus the $~$120k example training set in this paper.  Therefore, our recommendation for this problem domain and dataset size is to avoid using the function encoder (e.g. use BiLSTM).  

BiLSTM also performs significantly better overall than the baseline heuristic.  As expected, the recall scores for the baseline are quite high: an observation from examining the gold set is that most of the summaries start in the first few tokens, and the mean length of an annotation is 6.5 tokens, so a simple heuristic is likely to capture the correct words from a majority of comments.  However, the problem is precision.  While the ROUGE-1 and ROUGE-L precision is slightly lower in Tables~\ref{tab:rq6} and~\ref{tab:rq7}, these numbers are only for the comments containing summaries.  The baseline heuristic always creates a prediction, leading to zero scores for is-empty.  This is a problem because it will introduce a large amount of noise into the dataset -- 22\% of the gold set consists of comments with no summary.  This is reflected in the unitization score, which for the baseline is -0.43 to the unified controlled non-expert set and -0.52 to the gold set, compared to 0.36 for BiLSTM.  Note that the 0.36 unitization score for BiLSTM is comparable to the 0.35 unitization score of the expanded set to the gold set (RQ$_5$), which is reasonable considering that a large portion of the training data for BiLSTM is from the expanded set.

Another factor is that BiLSTM tends to do well in cases when it makes a prediction, but ROUGE scores are reduced by empty predictions.  For example, the ROUGE-1 and ROUGE-L F1 compared to the gold set are 88.89\% and 90.03\%, respectively, for cases in which BiLSTM makes a prediction (F1-NE scores in \Cref{tab:rq7}).  These scores highlight that most of the errors are concentrated in cases when BiLSTM does not make a prediction -- the predictions it does make have a high degree of accuracy, comparable to the similarity of the unified controlled non-expert set to the gold set (RQ$_3$).

\section{Conclusion \& Reproducibility}
\label{sec:reproducibility}

In this paper, we propose semi-automated and fully-automated procedures for extracting summary descriptions from unstructured function comments.  We demonstrated that non-experts from crowdsourcing platforms such as Amazon Mechanical Turk can in many cases achieve performance similar to that of experts, at greatly reduced cost.  Likewise, our fully-automated solution (BiLSTM) achieves strong performance, with most of its error concentrated in cases in which it makes no prediction, when in fact a summary exists.  These results have a direct application for documenting legacy code, but an even more important long term benefit lies in dataset generation for training automatic summary generation tools.

We have established that reasonable performance is achievable at a rate of USD\$0.05 per annotation via Mechanical Turk (our expanded set annotations), or even less if coupled with a fully-automated solution (BiLSTM).  Whereas agreement-based professional annotation of a repository of 1m function would likely cost up to US\$870k (Section~\ref{sec:annotategold}), annotation via crowdsourcing and trained neural models may cost US\$50k or less.  In the long run, given that research and proposal planning is often limited by dataset creation costs, in our view our findings have a direct benefit to the community in assisting this planning process.

To assist other researchers and encourage reproducibility, we provide all raw and processed data as well as scripts and model implementations in our online appendix:

\textbf{\url{https://github.com/NoPro2019/NoPro_2019}}

\section*{Acknowledgment}
This   work   is   supported   in   part   by   the NSF CCF-1452959 and CCF-1717607 grants.   Any  opinions,  findings,  and  conclusions expressed herein are the authors' and do not necessarily reflect those of the sponsors.

\bibliographystyle{IEEEtran}
\bibliography{main} 

\end{document}